\documentclass{emulateapj}

\newcommand{\ha}{H$\alpha$\,}
\newcommand{\hii}{H{\sc ii}\,}
\newcommand{\msun}{$M_{\sun}$\,}
\newcommand{\zsun}{Z$_{\sun}$\,}
\newcommand{\Msy}{$M_{\sun}\,{\rm yr}^{-1}$\,}
\newcommand{\ecms}{ergs\,cm$^{-2}$\,s$^{-1}$\,}

\shorttitle{Radio continuum spokes in the Cartwheel}
\shortauthors{Mayya et al.}

\begin{document}

\title
{The Detection of nonthermal radio continuum spokes and the study of star 
formation in the Cartwheel}

\author{Y. D. Mayya\altaffilmark{1}, D. Bizyaev\altaffilmark{2,3,4},
R. Romano\altaffilmark{1}, J. A. Garcia-Barreto\altaffilmark{5}, 
and E. I.  Vorobyov\altaffilmark{6}}


\altaffiltext{1}
{Instituto Nacional de Astrofisica, Optica y Electronica, Luis Enrique Erro 1, 
Tonantzintla, C.P. 72840, Puebla, Mexico}
\altaffiltext{2}{University of Texas at El Paso, El Paso, TX, 79968, USA }
\altaffiltext{3}{Sternberg Astronomical Institute, Moscow, 119899, Russia }
\altaffiltext{4}{Isaac Newton Institute of Chile, Moscow branch}
\altaffiltext{5}{IA-UNAM, Mexico City, D.F., Mexico }
\altaffiltext{6}{Institute of Physics, Stachki 194, Rostov-on-Don, Russia
and Department of Physics and Astronomy, University of Western Ontario,
London, Ontario, N6A 3K7, Canada }

\begin{abstract}
New sensitive Very Large Array 20~cm continuum observations of the 
Cartwheel, the prototypical collisional ring galaxy, were carried out with 
the principal aim of tracing supernova remnants that are expected to lie 
in the wake of the expanding ring and in the ring itself. We detect 
predominantly nonthermal radio continuum emission from regions associated
with 13 ring \hii\ complexes. The emission interior to the ring is confined 
to structures that resemble spokes of the wheel. The spokes start near bright 
\hii\ complexes, and extend to around 6\arcsec (4\,kpc) inward in the direction 
of the geometrical center of the ring. There is no apparent positional
coincidence between the radio continuum and optical spokes. Radial 
distribution of intensity along the spokes suggests that the past star 
formation rate (SFR) in the Cartwheel was much lower than the current SFR. 
New \ha\ observations were used to revise the current SFR in the Cartwheel. 
The revised value is 18\,\Msy, 
which is a factor of 4 lower than the value reported previously, but is in good 
agreement with the SFR estimated from far infrared luminosity. About 30\% of 
the observed 20~cm continuum nonthermal emission seems to originate in 
processes that are not related to star formation. Revised SFR in the Cartwheel 
is comparable to that in the rest of the ring galaxies. 
\end{abstract}

\keywords{galaxies: individual (\objectname{Cartwheel}) --- 
galaxies: interactions}

\section{Introduction}

Ring galaxies represent a class of colliding galaxies in which nearly 
symmetrical density waves are driven into a disk as a result of an almost
bulls-eye collision with a compact galaxy \citep{lyn76, the77}. 
Hydrodynamical simulations reproduce some of the basic features observed in 
ring galaxies, such as rings or crescent-shaped structures \citep{her93,ger96}. 
Expanding density waves trigger star formation (SF), and hence a star-forming 
ring is expected to form in ring galaxies. The star-forming ring propagates 
outward with time, leaving behind an evolved stellar population in its wake. 
Tracers of massive short-lived stars (e.g. \ha\ emission) are expected to 
delineate the present position of the wave, while tracers of the remnants of 
massive stars (e.g. nonthermal radio continuum) are expected to illuminate 
the regions in the wake of the expanding wave. Initial gas density 
distribution, and the velocity of the wave, would determine the shapes of the 
radial intensity profiles in the tracers of the young and old components 
\citep{kor01}.  

The Cartwheel is among the biggest known ring galaxies, in both angular and
linear size, and is considered to be the prototype of a ring galaxy. 
At a distance of 137~Mpc ($H_0$ = 65~km\,s$^{-1}$\,Mpc$^{-1}$), its angular
size of 72\arcsec\ corresponds to a linear diameter of 48~kpc. 
The entire population of bright \hii\ regions lies in the outer ring,
with more than 80\% of the current SF occurring in a few \hii\ complexes in the 
southern sector of the ring (Higdon 1995, hereafter H95). There are, however, a 
few faint regions inside \citep{amr98}. H95 derived a star formation rate (SFR) 
of 67~\Msy (159~\Msy for the H$_0$ adopted in this work), based on \ha\ flux, 
which is almost an order of magnitude higher than that derived from 
far infrared (FIR) flux.
Some of this discrepancy is probably due to problems in the H95 calibration
of \ha\ fluxes (J. L. Higdon 2000, private communication).

We have carried out sensitive 20~cm radio continuum (RC) observations using 
the Very Large Array (VLA\footnote{The VLA is part of
National Radio Astronomy Observatory, which is a facility of the National
Science Foundation operated under cooperate agreement by associated
Universities, Inc.}) with the goal of detecting nonthermal emission 
in the wake of the wave. We also carried out new \ha\ observations in order
to resolve the discrepancy between the SFRs derived from \ha, and FIR
luminosity. We present details of the new observations in Section~2. In 
Section~3, we describe the observed radio structures and also discuss the SFRs 
derived from \ha, FIR and RC fluxes. In Section~4, we discuss the results
obtained from these new observations from the context of expanding wave models.

\section{New RC and \ha\ Observations}

RC observations at 20 cm were carried out with the BnA configuration of the 
VLA during 2002 May and June, spread over seven runs totaling 35 hr
of integration time. The flux scale, antenna gain, and phase constants were 
calibrated using observations of standards 0137+331 (16.32~Jy at 1.38~GHz) 
and 0025$-$260 (8.70~Jy at 1.38~GHz). Data for each run were calibrated before 
combining them to produce the final map. The final cleaned and self-calibrated 
maps have almost a circular beam of FWHM 
of $3.3\arcsec\times3.0\arcsec$ (P.A.=$36^\circ$). Several 
maps with different weight parameters and de-convolved beam sizes
were obtained. Some maps were only cleaned and others were the result of
several iterations of self-calibration with different restoring beam sizes.
The rms or $\sigma$ noise of our final image is 30$\mu$Jy\,beam$^{-1}$, 
which is an improvement by around a factor of 4 over previous 
observations (Higdon 1996, hereafter H96). Although some of the radio 
structures that we discuss here are at the nominal 3\,$\sigma$ limit or just 
below it, we believe that these structures are true because of their presence 
in various trial maps we had produced during the reduction process.

New \ha\ CCD imaging observations were carried out during one photometric night
(2003, September 30) at the McDonald observatory (Texas) with a 
30$^{\prime\prime}$ telescope. We used three interference filters: 6700/123,
6560/100, and 6670/82, designated so for their central wavelengths and widths, 
both in units of \AA. The first one intercepts the redshifted \ha\ line from 
the galaxy (6761\,\AA), and the other two filters facilitate subtraction 
of the continuum flux entering the \ha\ filter. We took two exposures of 
the galaxy (totaling 35 minutes) as well as the standard stars in each 
filter. The exposures of the galaxy were bracketed in time and zenith distance 
by the calibration stars LTT\,377 and LTT\,1020 \citep{hamuy98}.

The measured total RC and \ha\ fluxes, along with errors on our measurements
are given in Table~1. Our 20~cm flux is in good agreement with the value
published by H96 (11.5 mJy). \ha\ fluxes in this table have been corrected 
for 10\% contribution from [NII] lines \citep{fosbury77} and galactic 
extinction (2.6\%). Our value is around 4 times lower than that given by H95. 

We used our new total flux to calibrate an excellent quality \ha\ image 
(seeing FWHM\,=\,0.8\arcsec) that is available at the Canada France Hawaii 
Telescope data archive. Fig.~1 shows the 20\,cm contours superposed 
on gray-scale representation of an \ha\ image that is smoothed to the 
resolution of the radio image, and full resolution {\it Hubble Space Telescope} 
($HST$) $B$-band image in panels ($a$) and ($b$) respectively. In the following 
section we give a description of the different observed features.

\begin{center}
\begin{deluxetable}{lccc}
\tablewidth{0pc}
\tablecaption{FIR, \ha and 20 cm fluxes and SFRs}
\tablehead{
\colhead{Quantity} & \colhead{FIR} & \colhead{\ha}
                & \colhead{20 cm} \\ 
}
\startdata
 Flux & $-$13.23 $\pm$0.12 & $-$12.40$\pm0.04$ & 12.8 $\pm$0.5 \\ 
         & $\log$ W\,m$^{-2}$   & $\log$ ergs\,s$^{-1}$\,cm$^{-2}$ & mJy \\ 
log Luminosity & 10.50 & 41.95 & 29.46 \\ 
         & L$\sun$    & erg\,s$^{-1}$ & W\,Hz$^{-1}$ \\ 
SFR (\Msy)  &       &      &       \\ 
\hspace{8mm} A$_{\rm v}=0.0$ & 17.71 & 7.91 & 29.13 \\ 
\hspace{8mm} A$_{\rm v}=1.2$ & 17.71 & 18.13 & 27.67 \\ 
\hspace{8mm} A$_{\rm v}=1.7$ & 17.71 & 25.60 & 26.61 \\ 
\hline
\enddata
\end{deluxetable}
\end{center}

\section{RC morphology and the nature of radio emission}

It can be seen in Figure~1 that the RC morphology of the Cartwheel resembles 
very much the morphology of the ring in \ha\ emission. Radio emission is 
detected from at least 13 \hii\ complexes. These complexes are among the 
brightest 15 in \ha\ emission (brighter than $8.31\times10^{-15}$~\ecms). 
The two non-detected sources (CW-6 and CW-29 of H95) are among the regions 
with the highest \ha\ equivalent widths, suggesting
that they are most likely too young to turn-on nonthermal RC emission.
Faint emission is also detected at the nuclear position (Fig.~1, {\it cross}). 
In the bright southern part of the ring, the RC image shows at least
seven finger-like structures that seem to originate from parts of the 
ring near bright \hii\ regions, and point radially inward to the geometrical 
center of the ring, giving the appearance of spokes of the wheel.
The most prominent spoke originates from close to CW-17, the brightest \hii\  
region in the Cartwheel. This is the only spoke that had been detected
in previous RC images (H96). The spoke originating near CW-20 crosses the 
ring, although with a slight change in the direction. Interestingly, the 
hyperluminous X-ray source reported by \cite{gao03} coincides with this region.

Are the RC spokes related to their optical counterparts? 
The first impression 
from Figure 1(b), is that the relation between the two, if any, is weak.
The RC spokes are short and straight, whereas the stellar spokes connect 
the two rings of the Cartwheel by a curved path.
The brightest RC spoke has no 
correspondence to any optical spoke even when extrapolated inward.
Of the other six RC spokes, only one (corresponding to CW-10), when extrapolated
inward meets an optical spoke. 
On the other hand, some of the weak RC emission interior to the main ring seems 
to coincide with the brightest points of the stellar spokes. 
So it is possible that the RC spokes are also longer and curved like their
optical counterparts, but we have detected only the bright part of that. 
Clearly, deeper RC images are required to address this question.

The azimuthal distribution of nonthermal RC intensities, and the thermal
fraction, in three successive radial zones, one corresponding to the 
position of the ring (radius=36\arcsec), and the other two positioned 
6\arcsec\ on either side of the ring, are plotted in Figures~2(a--c). 
Each zone has a width of 6\arcsec. The thermal RC flux is calculated from 
extinction corrected \ha\ flux using the relation given by \citet{condon92} 
for an electron temperature of 15000~K \citep{fosbury77}, and visual 
extinction A$_{\rm v}=$1.2~mag. 
The azimuthal RC intensity profile of the ring illustrates a well-known fact:
regions in the southern part of the ring (PA = 120--240$^\circ$) are 
several times brighter than elsewhere in the ring. The spokes stand out in
the azimuthal profile of the annular zone in the wake of the ring. 
The RC emission from the spokes is predominantly nonthermal.
The azimuthal profile exterior to the ring resembles that of the ring. 
However, unlike in the ring where 8--20\% of the observed RC flux of 
\hii\ complexes is of thermal origin, it is less than 2\% in the external 
zone except for the region just outside CW-12. 

\subsection{Global SFR from H$\alpha$, RC and FIR luminosities}

FIR, \ha\ and 20 cm RC fluxes have been extensively used as tracers of SFR in 
galaxies \citep{ken98}. The FIR flux, as defined in \citet{condon92}, is 
calculated for the Cartwheel using the IRAS 60 and 100$\mu$m fluxes from 
\citet{app87}. The resulting value is given in Table~1, along with the \ha\ 
and 20~cm RC flux measured in this work. SFRs are derived from the 
luminosities in FIR, \ha\ and 20~cm RC bands using the relations given by 
\citet{thronson86}, \citet{ken83}, and \citet{cy90}, respectively. 
Additive constants that convert log(luminosity) of row 2 to log(SFR) are 
 $-$9.27, $-$41.05, and $-$20.98 for FIR, \ha\ and 20~cm RC bands, 
respectively. The RC conversion factor depends on the nonthermal power-law 
index $\alpha$. A value of $\alpha=0.65$ is used in this work, which is based 
on the measurement for two of the brightest \hii\ regions in the 
Cartwheel (H96). In all these relations the initial mass function (IMF) 
from \citet{ken83} was used.
Use of Salpeter IMF would decrease the SFRs by 16\%  \citep{punuzzu04}.

In order to calculate the SFR from \ha\ fluxes, we need to correct for the 
internal extinction in \hii\ regions. Maximum value of A$_{\rm v}$ allowed
by the observed mean colors of the ring is 1.3~mag \citep{kor01}, as compared 
to A$_{\rm v}=2.1$~mag measured for the brightest \hii\ region. In the last 
three rows of Table~1, SFRs are tabulated for A$_{\rm v}=$0, 1.2 and 1.7 mag 
respectively. FIR-derived SFR is independent of extinction, where as the SFR 
derived from nonthermal RC flux is weakly dependent on extinction, which is 
due to the fact that it is obtained by subtracting the thermal component from 
the total, and that the thermal component is calculated from the 
extinction-corrected \ha flux. For A$_{\rm v}=$1.2\,mag, SFR derived from \ha\ 
agrees with that from FIR, whereas the RC gives around 30\% more SFR. The 
thermal contribution to the global RC flux is 10\%, which is in good agreement 
with the value found for star-forming galaxies \citep{condon92}. An SFR of 
18\,\Msy\ for the Cartwheel compares well with the SFRs of other collisionally 
formed ring galaxies and interacting galaxies \citep{mar95, dopita02}. 

RC emission of star-forming galaxies is correlated with the FIR emission.
Hence 30\% excess SFR derived from RC implies that 30\% of the detected
RC emission is not related to SF. What is the reason for this excess RC 
emission in the Cartwheel? A comparison of the RC and \ha\ morphologies 
suggests that the regions with excess RC emission are CW-20, CW-24 and the 
outer boundaries of these regions. Non-thermal radio emission not associated 
with SF has been discovered in shocked regions in a number of interacting 
galaxies, including in the intersection of the two merging galaxies forming 
Antennae \citep{mazza88}. Considering that the ring in the Cartwheel is 
expanding at high velocities, it is reasonable to think that the shocked gas 
contributes to a part of the observed nonthermal RC emission. However, 
published HI data of the Cartwheel, do not show evidence of high velocities 
suggestive of shocks in these regions. This implies that the neutral gas is
not coming from the same region as the shocked ionized gas. 

\section{Discussions}

The expanding density wave is expected to leave in its wake old stars and 
remnants of massive stars. Non-thermal RC emission originates in supernova 
remnants, which trace stars that are formed $\sim$10--50~Myr ago 
(lifetime of stars more massive than 9~\msun\ at Z=\zsun/10). Hence 
nonthermal RC emission from spokes represent the trajectory of the 
star-forming wave over the past $\sim$50~Myr. Intensity distribution along 
the spokes has valuable information on the past SF process, and can be
used to constrain the models of SF in an expanding density wave. In 
Figure~3(a), we plot the radial profiles of RC and \ha\ intensities expected 
from models of SF in an expanding density wave. In particular, we used the 
\citet{vb01} model, in which the density wave is assumed to be of Gaussian 
shape $\sigma=1$~kpc, propagating at 90~km~s$^{-1}$ in an exponential gas 
disk of scale-length of 20~kpc. A Schmidt law with power law index n=1.5 is 
used to scale gas density to SFR. The RC emission, which traces stars 
formed $\sim$50~Myr ago, lags behind the \ha\ peak, which marks the current 
location of the wave, by about 2\arcsec. We have carried out a series of 
tests, and confirmed that the predicted displacements in the peak positions 
of the profiles should be seen even at the spatial resolution of our RC image.

We obtained radial intensity cuts on the \ha\ and RC images, passing through 
the ring center (marked by a small circle in Figure~1) and the RC detected 
parts of the ring. All the resulting 
profiles resemble one of the three representative profiles of Figure~3(b-d).
The majority of the RC profiles are skewed inward as compared to the \ha\ 
profile, with the peaks of the two profiles coinciding as in (b). 
In some directions, the RC emission peaks outside the \ha\ profile (c).
RC and \ha\ profiles of the brightest \hii\ complex (CW-17) are identical 
except that the former is broader. Surprisingly, none of the regions 
show the expected inward shift of the RC profile with respect to \ha\ profile. 
The profiles skewed inward correspond to the zones where spokes are 
detected in the wake of the wave, where as profiles skewed outward are 
basically zones where radio emission extends outside the \ha\ ring. As
discussed in the previous section, shocks ahead of the wave are responsible
for the RC emission in these zones.

In the Cartwheel, the $K$-band intensity profile, which traces the red 
supergiant stars that are formed about 10--20~Myr ago, is found to peak
marginally inside the \ha\ profile \citep{vb03}. The displacements, are 
consistent with a wave velocity of $\approx100$~km~s$^{-1}$. The displacements 
were not noted along the minor axis cuts, which \citet{vb03} explained as due 
to orientation effects. \hii\ regions CW-6 and CW-28 are along the major
axis, both of which contain radially directed RC spokes, but the expected
2\arcsec\ shift in the position of peak of the RC emission cannot be noted.

The lack of radial separation between the RC and \ha\ profiles calls for a 
revision of the physical ingredients that go into the models, especially in the
treatment of SF and feedback effects. The observed profile shapes imply 
that the SFR increased to the currently observed values abruptly a few tens 
of Myr ago. A possible reason for the lack of SF in the past could be that
the pre-collisional gas disk was well below criticality and the density 
enhancements caused by the passing wave were not sufficient to drive the 
densities above the critical value required for collapse. 
Even if the disk was critical before the interaction, the passing
wave can heat the disk, making it subcritical. Hence realistic ring models
should include heating and cooling of the disk. Observations of more ring
galaxies in the RC would be invaluable in working out the details of the 
physics of SF in an expanding density wave.

\acknowledgments
We thank the staff at NRAO, especially Barry Clark, for liberal allotment of 
VLA time. We also thank the staff at
McDonald observatory, and personally David Doss. This work has gained from
discussions with Alessandro Bressan and Vladimir Korchagin at various stages of 
the project. We thank the anonymous referee whose comments helped to improve 
the presentation of the manuscript. This work was supported by research
grants 39714-F (CONACyT, Mexico), 99-04-OSS-058, and 04-02-16518 (JPL/NASA and 
RFBR).


\clearpage

\begin{figure} 
\epsscale{1.15}
\plottwo{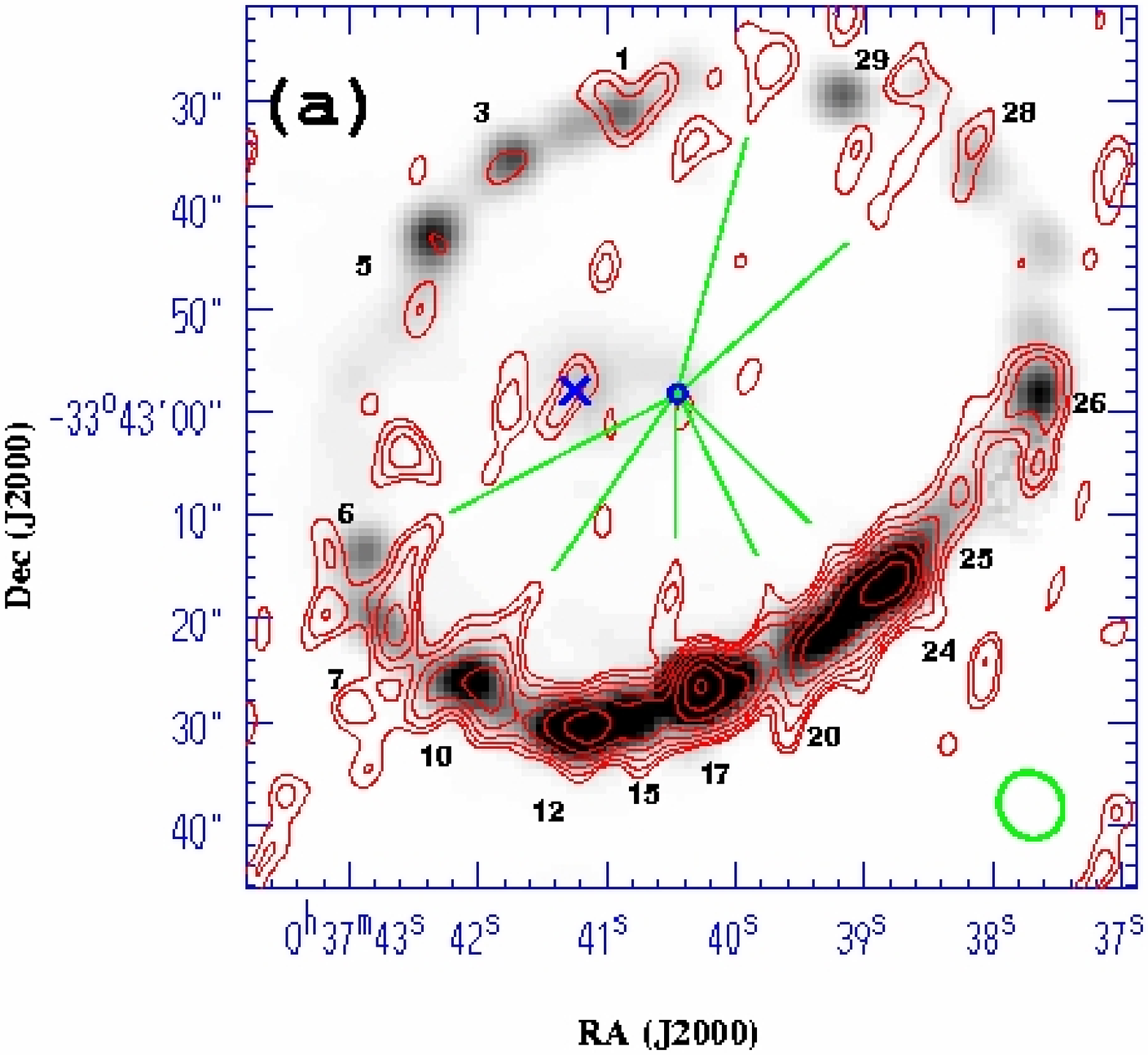}{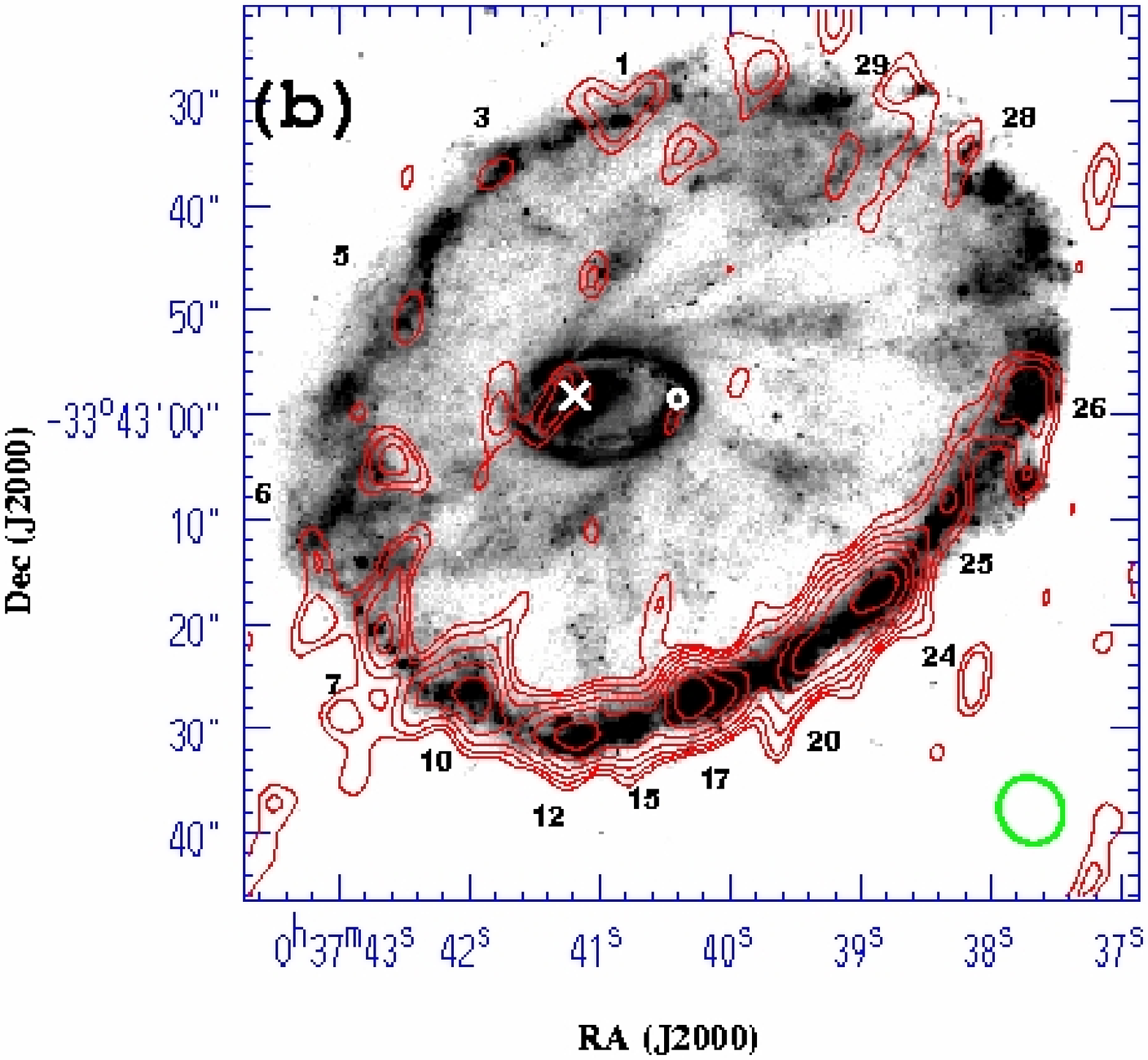}
\caption{20 cm RC intensity contours of the Cartwheel superimposed on (a)
an \ha\ image (gray scale), which has been smoothed to the resolution of the
20~cm RC image, and (b) the HST B-band image. The lowest contour level 
corresponds to 60\,$\mu$Jy\,beam$^{-1}$ ($\approx2\sigma$), and the subsequent
contour levels increase by a factor of $\sqrt{2}$. The ellipse at the 
bottom-right corner indicates the RC beam size. In (a) note the excellent positional 
correspondence between radio peaks and \hii\ complexes, which have been 
labeled by their H95 numbers. Straight lines are drawn connecting the 
filamentary structures or {\it spokes} to the geometrical center of the ring. 
Unlike the optical spokes (features connecting the inner ring to the outer ring
in (b)), the RC spokes are straight and short. 
The position of the nucleus is marked by a cross.
}
\end{figure}
%
\begin{figure}[htb]
\epsscale{1.25}
\plotone{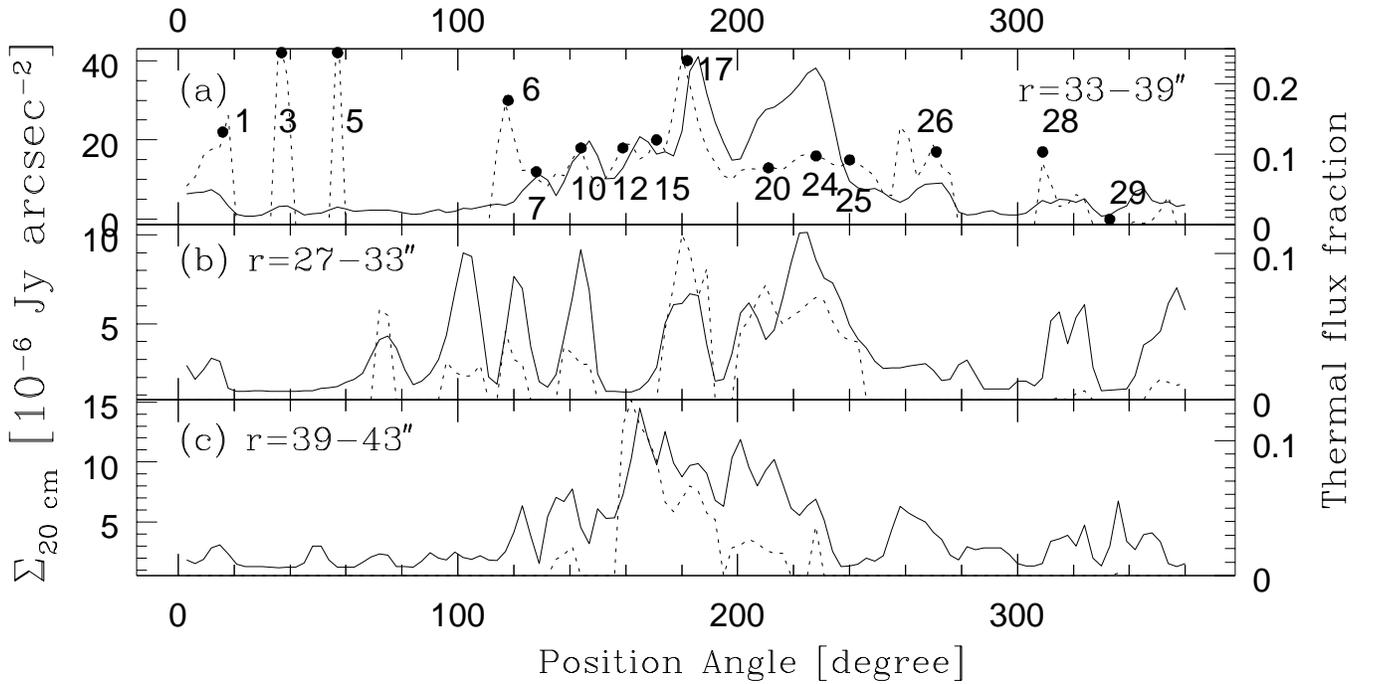}
\vspace*{-7.5cm}
\caption{Azimuthal profiles of nonthermal radio continuum emission 
(solid line) (a) in the ring (b) inside the ring and (c) outside the ring.
The thermal fraction at 20~cm is plotted by the dotted line.
The brightest 15 \hii\ regions are labeled. Radio continuum spokes are clearly
seen in (b) at position angles 120, 145, 180, 205, 225, 320 and 355 degree.
}
\end{figure}
%
%
\begin{figure}
\epsscale{1.20}
\plotone{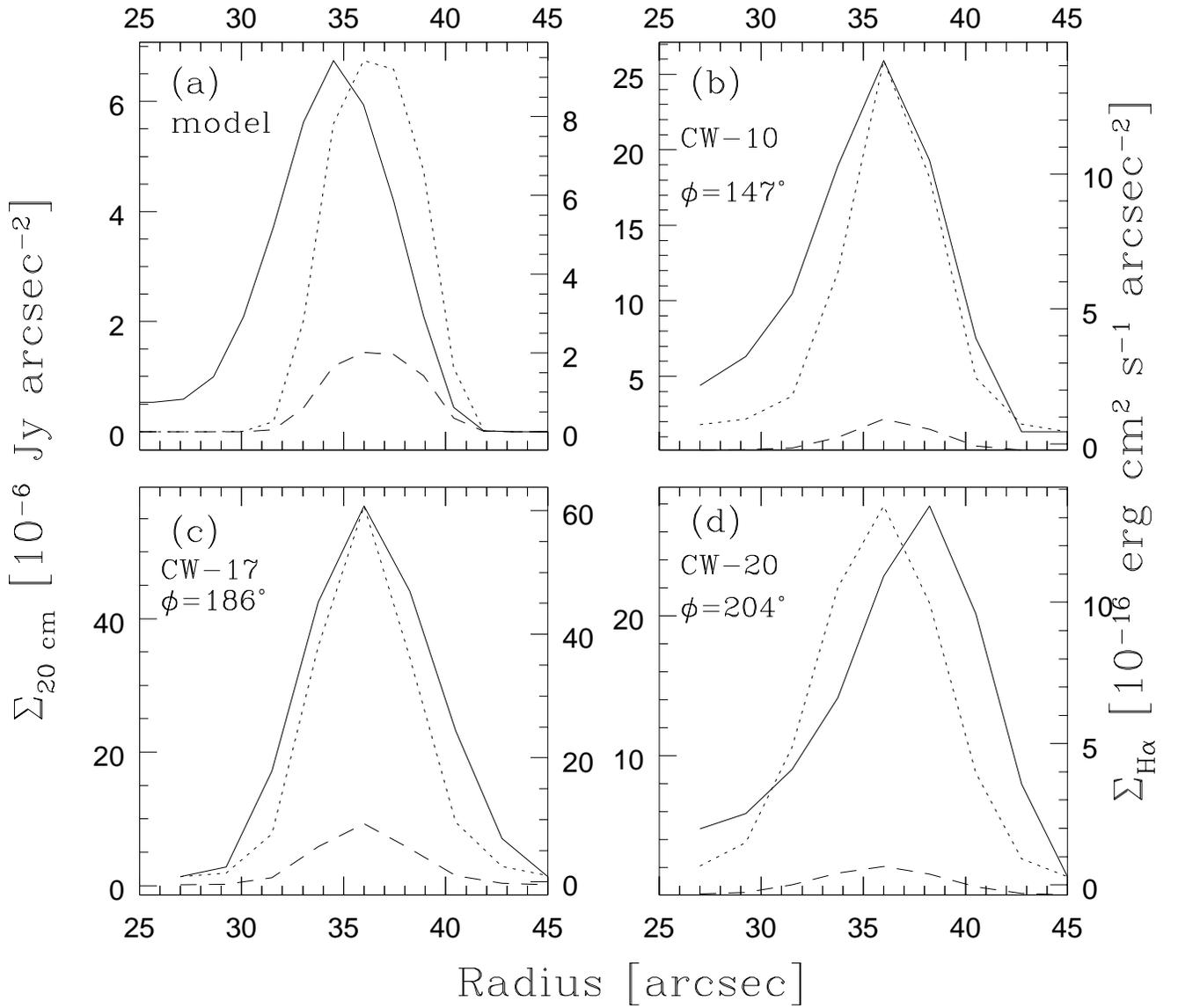}
\vspace*{-1.5cm}
\caption{The radial intensity cuts along 3 representative \hii\ complexes
compared to a model. The solid lines represent the RC intensities (scale on the 
left), where as the dotted lines (scale on the right) denote the intensity 
of the \ha\ emission. The dashed lines correspond to the estimated 20~cm 
thermal component. A$_{\rm v}=1.2$~mag is assumed. The H95
identification of the \hii\ regions, as well as the position
angle ($\phi$) of the regions are denoted in each panel. Each profile
is obtained by averaging in wedges of 3$^\circ$ opening angle. 
}
\end{figure}


\begin{thebibliography}{}
%
\bibitem[Amram et al. (1998)]{amr98}
 Amram, P., Mendes de Oliveira, C., Boulesteix, J., \& Balkowski, C. 1998,
 \aap, 330, 881 
%
\bibitem[Appleton \& Struck-Marcell(1987)]{app87}
 Appleton, P. N., \& Struck-Marcell, C. 1987, \apj, 312, 566
%
\bibitem[Condon (1992)]{condon92}
Condon, J. J.  1992, \araa, 30, 575
%
\bibitem[Condon \& Yin (1990)]{cy90}
Condon, J. J., \& Yin, Q. F. 1990, \apj, 357, 97
%
\bibitem[Dopita et al.  (2002)]{dopita02}
Dopita, M. A., Pereira, M., Kewley, L. J., \& Capaccioli, M. 2002, \apjs, 143, 47
%
\bibitem[Fosbury \& Hawarden (1977)]{fosbury77}
 Fosbury, R. A. E., \& Hawarden, T. G. 1977, \mnras, 178, 473
%
\bibitem[Gao et al. (2003)]{gao03}
Gao, Y., Wang, Q. D., Appleton, P. N., \& Lucas, R. A. 2003, \apjl, 596, 171
%
\bibitem[Gerber, Lamb \& Balsara(1996)]{ger96}
 Gerber, R. A., Lamb, S. A., \& Balsara, D. S. 1996, \mnras, 278, 345
%
\bibitem[Hamuy et al. (1998)]{hamuy98}
Hamuy et al. 1994, \pasp, 106, 566 
%
\bibitem[Hernquist \& Weil(1993)]{her93}
 Hernquist, L., \& Weil, M. L. 1993, \mnras, 261, 804
%
\bibitem[Higdon(1995)]{hig95}
 Higdon, J. L. 1995, \apj, 455, 524 (H95)
%
\bibitem[Higdon(1996)]{hig96}
 Higdon, J. L. 1996, \apj, 467, 241 (H96)
%
\bibitem[Kennicutt (1983)]{ken83}
Kennicutt, R. C. 1983, \apj, 272, 54
%
\bibitem[Kennicutt(1998)]{ken98}
Kennicutt, R. C. 1998, \araa, 36, 189
%
\bibitem[Korchagin, Mayya \& Vorobyov (2001)]{kor01}
Korchagin, V., Mayya, Y. D., \& Vorobyov, E. I. 2001, \apj, 554, 281
%
\bibitem[Lynds \& Toomre(1976)]{lyn76}
Lynds, R., \& Toomre, A. 1976, \apj, 209, 382
%
\bibitem[Marston \& Appleton (1995)]{mar95}
Marston, A. P., \& Appleton, P. N. 1995, \aj, 109, 1002

\bibitem[Mazzarella et al. (1988)]{mazza88}
Mazzarella, J. M., Gaume, R. A.. Aller, H. D., \& Hughes, P. A. 1988, \apj, 333, 168
%
\bibitem[Punuzzu et al. (2004)]{punuzzu04}
Punuzzu, P., Bressan, A., Granato, G. L., Silva, L., \& Danese, L. 2004, 
\aap, 409, 99
%
\bibitem[Theys \& Spiegel(1977)]{the77}
Theys, J. C., \& Spiegel, E. A. 1977, \apj, 212, 616
%
\bibitem[Thronson \& Telesco (1986)]{thronson86}
Thronson, H. A., \& Telesco, C. M. 1986, \apj, 311, 98
%
\bibitem[Vorobyov \& Bizyaev (2001)]{vb01}
Vorobyov, E. I., \& Bizyaev, D. 2001, \aap, 377, 835
%
\bibitem[Vorobyov \& Bizyaev (2003)]{vb03}
Vorobyov, E. I., \& Bizyaev, D. 2003, \aap, 400, 81
%
\end{thebibliography}
\end{document}